\documentclass[a4paper,11pt]{article}
\pdfoutput=1

\usepackage{a4wide}
\usepackage[latin1]{inputenc}
\usepackage{dsfont}
\usepackage{amsfonts}
\usepackage{amsmath}
\usepackage{amssymb}
\usepackage{amsthm}
\usepackage{graphicx}
\allowdisplaybreaks[1]
\usepackage{color}
\usepackage{ulem}
\usepackage{multirow}

\usepackage[small,bf]{caption}
\newcommand{\braket}[1]{\langle #1 \rangle}

\usepackage{commath}
\usepackage{mathcomp}
\usepackage{colortbl}
\usepackage{bbold}
\usepackage{commath}

\usepackage{mathtools}

\newcommand{\Hubble}{ \mathcal{H} }

\newcommand{\kahler}{K\"{a}hler }
\newcommand{\kahlerdriven}{K\"{a}hler-driven }
\usepackage{slashed}

\newcommand{\mpl}{m_{\rm Pl}}

    \usepackage{array}
    \newcolumntype{C}[1]{>{\centering\arraybackslash}p{#1}}

\begin{document}

\begin{titlepage}

\vspace*{-15mm}
\vspace*{0.7cm}

\begin{center}

{\Large {\bf Realising effective theories of tribrid inflation:\\Are there effects from messenger fields?}}\\[8mm]

Stefan Antusch$^{\star\dagger}$\footnote{Email: \texttt{stefan.antusch@unibas.ch}} and
David Nolde$^{\star}$\footnote{Email: \texttt{david.nolde@unibas.ch}}

\end{center}

\vspace*{0.20cm}

\centerline{$^{\star}$ \it
Department of Physics, University of Basel,}
\centerline{\it
Klingelbergstr.\ 82, CH-4056 Basel, Switzerland}

\vspace*{0.4cm}

\centerline{$^{\dagger}$ \it
Max-Planck-Institut f\"ur Physik (Werner-Heisenberg-Institut),}
\centerline{\it
F\"ohringer Ring 6, D-80805 M\"unchen, Germany}

\vspace*{1.2cm}

\begin{abstract}
\noindent
Tribrid inflation is a variant of supersymmetric hybrid inflation in which the inflaton is a matter field (which can be charged under gauge symmetries) and inflation ends by a GUT-scale phase transition of a waterfall field. These features make tribrid inflation a promising framework for realising inflation with particularly close connections to particle physics. Superpotentials of tribrid inflation involve effective operators suppressed by some cutoff scale, which is often taken as the Planck scale. However, these operators may also be generated by integrating out messenger superfields with masses below the Planck scale, which is in fact quite common in GUT and/or flavour models. The values of the inflaton field during inflation can then lie above this mass scale, which means that for reliably calculating the model predictions one has to go beyond the effective theory description. We therefore discuss realisations of effective theories of tribrid inflation and specify in which cases effects from the messenger fields are expected, and under which conditions they can safely be neglected. In particular, we point out how to construct realisations where, despite the fact that the inflaton field values are above the messenger mass scale, the predictions for the observables are (to a good approximation) identical to the ones calculated in the effective theory treatment where the messenger mass scale is identified with the (apparent) cutoff scale.
\end{abstract}
\end{titlepage}

\newpage

\section{Introduction}

Cosmological slow-roll inflation has proven to be a very successful paradigm in modern cosmology. Not only does it solve the horizon and flatness problems of homogeneous cosmology, it also correctly predicts the adiabatic, Gaussian and nearly scale-invariant primordial curvature perturbations in excellent agreement with the increasingly precise observations of the cosmic microwave background \cite{Ade:2015oja,Ade:2015tva} and large scale structure \cite{Beutler:2011hx,Anderson:2013zyy,Ross:2014qpa}.

One interesting class of models is tribrid inflation, a variant of supersymmetric hybrid inflation \cite{Dvali:1994ms,Linde:1997sj,Pallis:2009pq,Nakayama:2010xf}. In tribrid inflation \cite{Antusch:2004hd,Antusch:2008pn,Antusch:2009ef,Antusch:2010mv,Antusch:2013toa}, inflation is driven by a slow-rolling inflaton field (which can be charged under symmetries, including gauge symmetries \cite{Antusch:2010va}) and terminated by a GUT-scale particle physics phase transition in some waterfall field. These features make tribrid inflation a very promising framework for realising inflation with close connections to particle physics.

Like in many other models of inflation, non-renormalizable operators play an important role in tribrid inflation \cite{Antusch:2012jc}. For tribrid inflation, such operators are an essential part of the superpotential, whereas in other models they can provide important corrections. Non-renormalizable operators generally arise in the low-energy effective field theory (EFT) from integrating out physics at some higher energy scale $\Lambda_{\rm NP}$. It is usually assumed that $\Lambda_{\rm NP} \lesssim \mpl$, as we expect new physics to appear at least at the Planck scale, but $\Lambda_{\rm NP}$ can also be much smaller if there is some new physics between the electroweak and the Planck scale. For example, some heavy ``messenger'' particles of mass $m_A \ll \mpl$ would generate operators suppressed by $\Lambda_{\rm NP} \sim m_A$.

In general, an EFT is only valid for energies below the cutoff scale $\Lambda_{\rm cutoff} \sim \Lambda_{\rm NP}$. However, it is not immediately clear how this translates into bounds on the inflaton field displacement $\Delta \phi$ during inflation. For $\Delta \phi \gtrsim \Lambda_{\rm cutoff}$ it is not possible to deduce the most relevant operators by a truncated expansion in $\Delta \phi / \Lambda_{\rm cutoff}$. However, if there are symmetry arguments to limit the number and field content of non-renormalizable operators, the EFT might still work for larger $\Delta \phi$ as long as the relevant energy scale (e.g.\ the Hubble scale $\mathcal{H}$ in inflation) is smaller than $\Lambda_{\rm cutoff}$.

In tribrid inflation, this question is particularly relevant, as its waterfall phase transition can naturally be related to some new physics at high but sub-Planckian scales $\Lambda_{\rm NP} \ll \mpl$. At the same time, typical inflaton field displacements $\Delta \phi$ during tribrid inflation are about $M_{\rm GUT} \lesssim \Delta \phi \ll \mpl$, and many potentially interesting tribrid models have $\Delta \phi \gtrsim \Lambda_{\rm NP}$. It is therefore important to understand whether tribrid inflation in such models can be studied in the effective field theory framework with non-renormalizable operators suppressed by $\Lambda_{\rm cutoff} \sim \Lambda_{\rm NP}$, or if one needs to explicitly include all relevant particles with masses up to almost $\mpl$.

In this paper, we study this question for tribrid inflation by comparing the predictions of a non-re\-nor\-ma\-li\-za\-ble ``K\"{a}hler-driven'' tribrid model to an explicit UV completion in which the effective superpotential operators are replaced by renormalizable couplings to heavy messenger fields. We discuss one particular case in detail and show that the tree-level quantities match even for $\Delta \phi > \Lambda_{\rm cutoff}$ up to small corrections of order $\Hubble/\Lambda_{\rm cutoff}$. We then analyse the one-loop corrections which can be different between effective and renormalizable superpotential, and finally discuss how one can generate different effective tribrid models by different choices for the messenger sector, providing guidelines for tribrid inflation model building.

\section{Tribrid inflation}
\label{sec:tribrid}

In this paper, we will discuss tribrid inflation in supergravity with the superpotential \cite{Antusch:2013toa}
\begin{align}
 W_{\rm eff} \, &= \, S\left( \frac{H^4}{\Lambda_H^2} - \Lambda^2 \right) + \frac{1}{\Lambda_\phi} H \Phi^2 N + ..., \label{eq:effectiveW}
\end{align}
where the dots denote terms which are irrelevant for inflation.\footnote{Such extra terms can give a mass to $N$ after inflation, e.g.\ via $\Delta W \propto H^3 N^2$ if one chooses a $\mathbf{Z}_{16}$ symmetry ($n=4$) in table~\ref{tab:symmetryCharges}.} Our analysis can be extended to more general superpotentials, which we will briefly discuss in section~\ref{sec:generalisation}. However, the analysis will be much easier to follow for a particular example, and therefore we will do most of our analysis for the explicit superpotential in eq.~\eqref{eq:effectiveW}. The given form of the superpotential can be enforced e.g.\ by an $U(1)_R$ and a $\mathbf{Z}_{4n}$ symmetry with charge assignments given in table~\ref{tab:symmetryCharges}.

\begin{table}[bt]
\centering
\begin{tabular}{ | c | C{0.92cm} C{0.92cm} C{0.92cm} C{0.92cm} | }
  \hline
  & $S$ & $H$ & $N$ & $\Phi$ \\
  \hline
  $U(1)_R$ & $2$ & $0$ & $1$ & $1/2$ \\
  $\mathbf{Z}_{4n}$ & $0$ & $n$ & $n-2$ & $n+1$ \\
  \hline
\end{tabular}
\caption{One possible set of symmetries and charge assignments for the superpotential in eq.~\eqref{eq:effectiveW}, for any integer $n \geq 3$.}
\label{tab:symmetryCharges}
\end{table}

We also assume that the K\"{a}hler potential can be expanded in the modulus squared of the fields:\footnote{The given charge assignment allows for some other operators as well, like $\Delta K \propto \operatorname{Re}(H^4)$, but those do not have a significant effect in tribrid inflation models.}
\begin{align}
 K \, &= \, \sum_i \lvert Y_i \rvert^2 + \sum_{i j} \frac{ \kappa_{ij} }{ \mpl^2 } \lvert Y_i \rvert^2 \lvert Y_j \rvert^2 + \sum_{ijk} \frac{\kappa_{ijk}}{\mpl^4} \lvert Y_i \rvert^2 \lvert Y_j \rvert^2 \lvert Y_k \rvert^2 + ..., \label{eq:K}
\end{align}
with $Y = \left\{ S, H, N, \Phi \right\}$. We generally assume that the higher-order operators in $K$ are Planck-suppressed, i.e.\ $\kappa_{ij}, \kappa_{ijk} \lesssim \mathcal{O}(1)$.

\begin{figure}[tb]
  \centering
  \includegraphics[width=10cm]{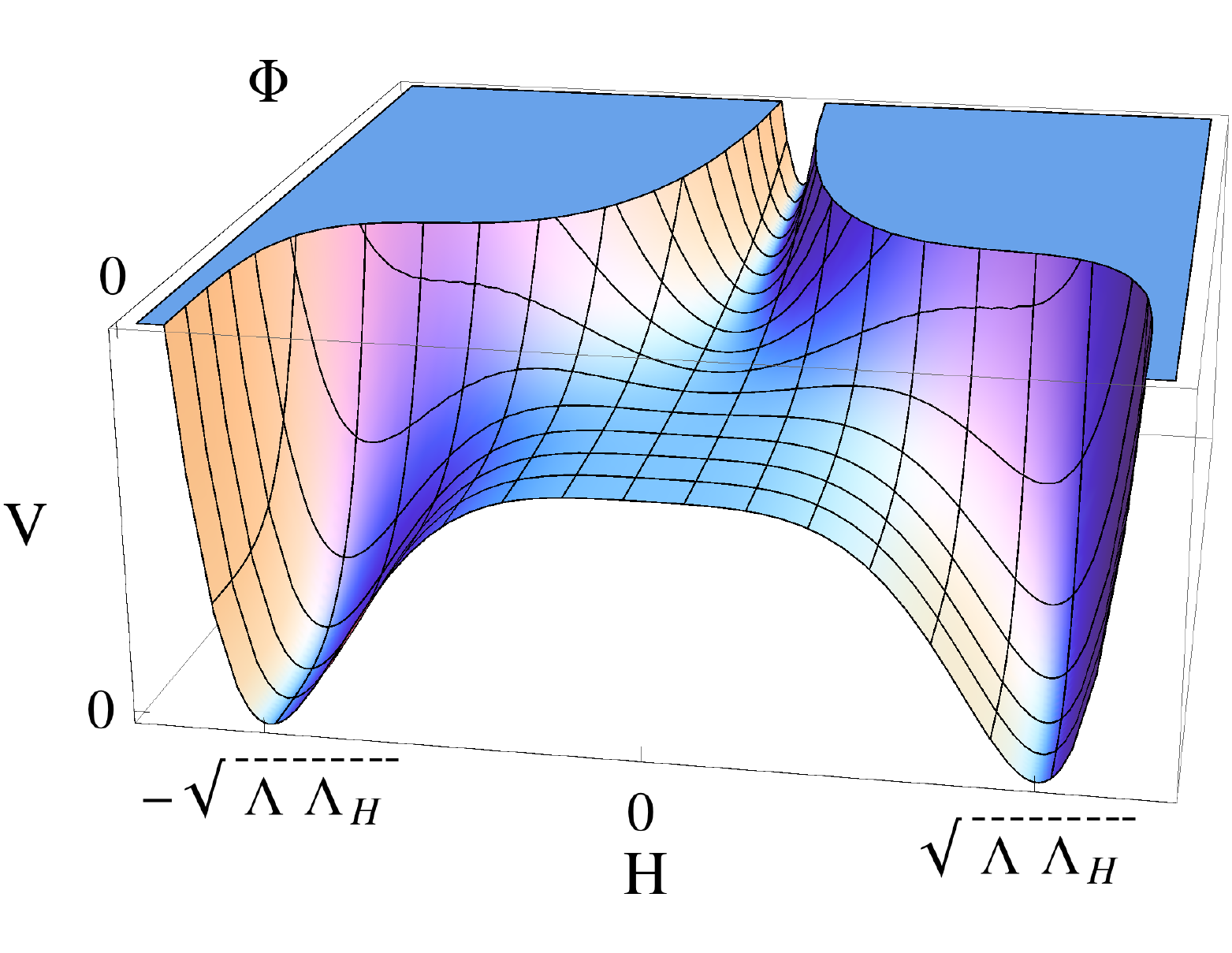}
  \caption{Example scalar potential in tribrid inflation. During inflation, the field slowly rolls down the nearly flat valley at $H=0$ from large $\Phi$ towards $\Phi \rightarrow 0$. While the field is slowly rolling, the large vacuum energy $V \simeq \Lambda^4$ drives inflation. After $\Phi$ passes the critical point $\Phi_c$, the waterfall field $H$ gets a tachyonic mass, and the field quickly rolls down the large slope towards the global minimum at $H^2 = \pm \Lambda_H\Lambda$ and $\Phi = 0$. In the global minimum, $V=0$ and inflation has ended.}
  \label{fig:hybridPotential3D}
\end{figure}

The resulting scalar potential is plotted in fig.~\ref{fig:hybridPotential3D}. It has the form \cite{Antusch:2012jc}
\begin{align}
 V \, &= \, \Lambda^4 \left| \frac{H^4}{\Lambda_H^2 \Lambda^2} - 1 \right|^2 \, + \, \left(  \frac{ \lvert \Phi \rvert^4 }{\Lambda_\phi^2} + \Delta m_H^2   \right)\lvert H \rvert^2  \, + \Delta V_{\rm inf}( \Phi ) \notag\\
 &~~~\quad+ \, \left(  \frac{ \lvert \Phi \rvert^4 }{\Lambda_\phi^2} + \Delta m_N^2   \right)\lvert N \rvert^2  \, + \, \Delta m_S^2 \lvert S \rvert^2 \, + ...,
\end{align}
where $\Lambda^4 + \Delta V_{\rm inf}(\Phi)$ is the inflaton potential during inflation and $\Delta m_H^2$, $\Delta m_N^2$ and $\Delta m_S^2$ are additional squared masses for $H$, $N$ and $S$ due to K\"{a}hler potential operators for which we assume $\Delta m_H^2 \lesssim - \Hubble^2$ and $\Delta m_N^2$, $\Delta m_S^2 \gtrsim \Hubble^2$.\footnote{The sign of $\Delta m_{Y_i}^2$ depends on $\kappa_{iS}$. We assume that $\kappa_{iS}$ is chosen such that tribrid inflation is possible; for $\Delta m_S^2 < 0$, $S=0$ would not be stable during inflation, and for $\Delta m_H^2 > 0$, $H$ would not develop a tachyonic instability for any $\phi$.}

Inflation happens for $\lvert \Phi \rvert > \lvert \Phi_{\rm c} \rvert \equiv \lvert \Delta m_H \Lambda_\phi \rvert^{1/2}$ and $H = N = S = 0$, with the inflaton slowly rolling towards smaller values while $H = N = S = 0$ due to their positive mass terms, and the universe expands due the large false vacuum energy $V \simeq \Lambda^4$. Eventually, $\lvert \Phi \rvert$ drops below $\lvert \Phi_{\rm c} \rvert$, at which time $H$ develops a tachyonic mass and quickly falls towards its minimum, terminating inflation.

It turns out that during inflation the model is equivalent to single-field inflation with the canonically normalized\footnote{The small difference between $\phi$ and $\sqrt{2}\lvert\Phi\rvert$ is due to canonical normalization, because $\Phi$ has a non-canonical kinetic term $K_{\overline{\Phi}\Phi}(\partial_\mu \Phi)^\dagger(\partial^\mu \Phi)$.} real inflaton field $\phi \simeq \sqrt{2}\lvert \Phi \rvert$ and the inflaton potential \cite{Antusch:2012jc}
\begin{align}
 V_{\rm inf} \, &= \, \Lambda^4 \left[ \frac{ e^{K/\mpl^2} }{ K_{\overline{S}S} } \right]_{S=H=N=0} \, \simeq \, \Lambda^4 \left(  1 \, + \, \frac{a}{\mpl^2} \left| \phi \right|^2 \, + \, \frac{b}{\mpl^4} \left| \phi \right|^4 \, + \, ...  \right), \label{eq:kahlerSimple1}
\end{align}
with
\begin{subequations}
\begin{align}
 \label{eq:kahlerVphiA}
 a \, &= \, \frac{1}{2} \left( 1 - \kappa_{S\Phi} \right), \\
 \label{eq:kahlerVphiB}
 b \, &= \, \frac18 +  a^{2} - \frac{a}{2} +  \kappa_{\Phi\Phi}\left( \frac14 - \frac{2a}{3} \right) - \frac{ \kappa_{S\Phi\Phi} }{4},
\end{align}
\end{subequations}
which ends by a waterfall transition at
\begin{align}
 \phi_{\rm c} \, = \, \sqrt{ \lvert 2\Lambda_\phi \Delta m_H \rvert }, \label{eq:kahlerSimple2}
\end{align}
after which the waterfall field acquires a vacuum expectation value
\begin{align}
 \braket{H^2} \, = \, \pm \Lambda_H \Lambda.
\label{eq:kahlerSimple3}
\end{align}

The CMB predictions for this model have been derived in \cite{Antusch:2012jc}. For $\phi \ll \mpl$, which is required for the expansion in eq.~\eqref{eq:K}, it generally predicts $\alpha_s \gtrsim 0$ and $r \lesssim 0.01$, as well as relations between $\alpha_s$, $\Lambda$, $\Lambda_\phi$ and $\phi_c$. $n_s$ can take any value in the range allowed by CMB observations. These predictions are based only on eqs.~\eqref{eq:kahlerSimple1}, \eqref{eq:kahlerSimple2} and \eqref{eq:kahlerSimple3}, and any other model that can be reduced to these three equations must lead to identical predictions.

\section{Generating $W$ from renormalizable couplings}
\label{sec:renormalizable}

In the tribrid inflation model discussed above, the inflaton field traverses distances $\Delta \phi \sim \mathcal{O}(\mpl/10)$ during inflation. While this is sufficiently small to allow expanding the potential in powers of $\phi^2/\mpl^2$, it is not clear whether the model remains valid for $\Lambda_H \ll \mpl$ or $\Lambda_\phi \ll \mpl$: this would lead to $\Delta \phi \gtrsim \Lambda_H$ or $\Delta \phi \gtrsim \Lambda_\phi$, which suggests that the inflaton field value might be above the cutoff scale of the EFT during inflation.

To study this question, we construct an explicit UV completion of the above model by replacing the non-renormalizable superpotential operators with renormalizable couplings to heavy messenger fields $A_i$ and $B_i$ which have masses much greater than the Hubble scale.\footnote{If the messenger fields' masses are below the Hubble scale, they need to be taken into account as dynamical degrees of freedom. Their quantum fluctuations can affect the primordial spectrum of perturbations, and the predictions must be calculated using a multi-field formalism such as the $\delta N$ formalism.} We then study inflation with the renormalizable superpotential and show that it leads to the same tree-level prediction as the EFT even for $\Delta \phi > \Lambda_H$ and $\Delta \phi > \Lambda_\phi$, up to small corrections. The one-loop quantum corrections to the inflaton potential and the effects of different choices for the messenger sector will be discussed in sections \ref{sec:loop} and \ref{sec:generalisation}, respectively.

\subsection{$W$ and $K$ with renormalizable couplings to messenger fields}

\begin{figure}[bt]
  \centering$
\begin{array}{ccc}
\includegraphics[height=0.166\textwidth]{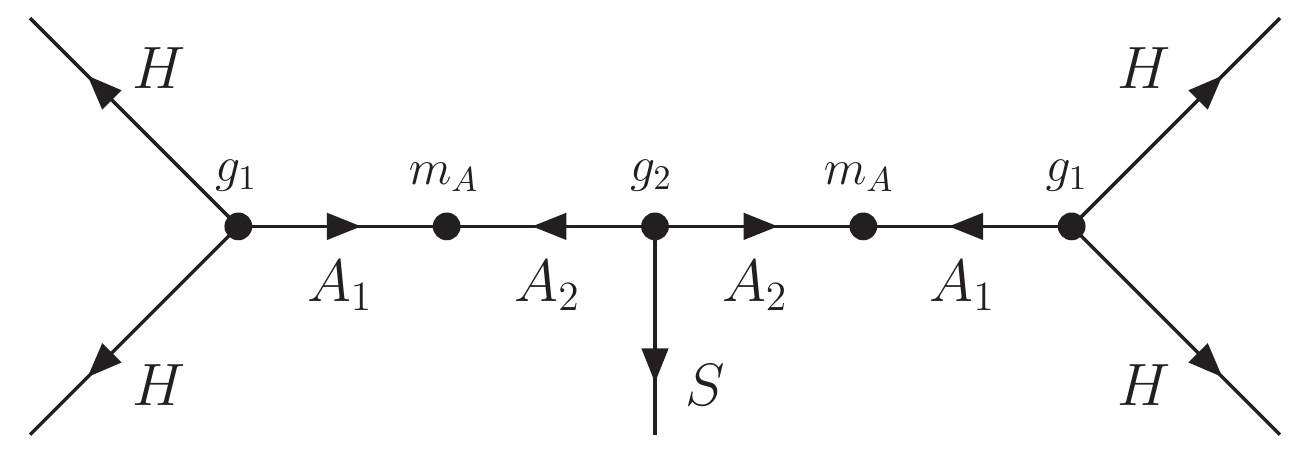} & \hphantom{.....} &
\includegraphics[height=0.166\textwidth]{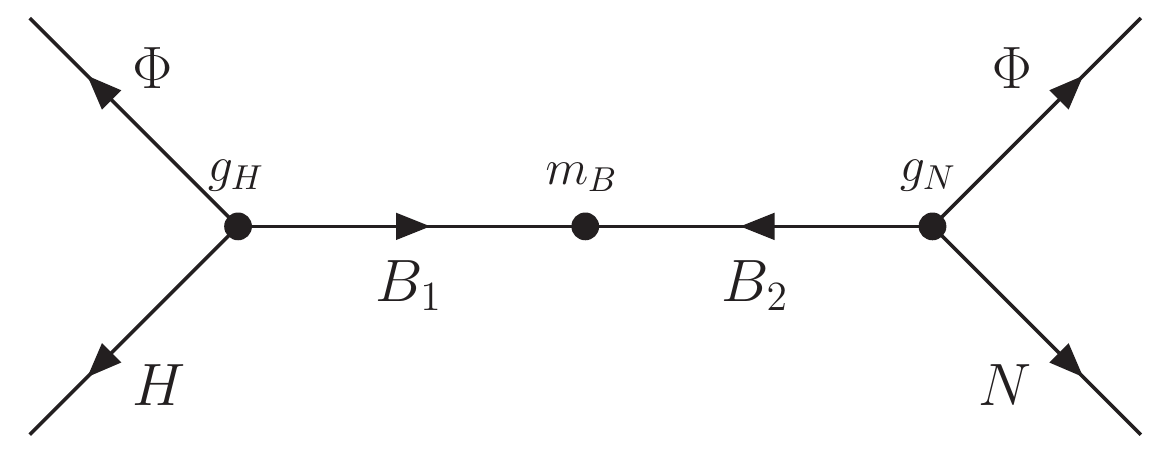}
\end{array}$
  \caption{Diagrams for generating the non-renormalizable operators of the superpotential in eq.~\eqref{eq:effectiveW} from renormalizable superpotential couplings to heavy messenger fields $A_i$ and $B_i$.}
  \label{fig:messengerDiagrams}
\end{figure}

We can generate the two non-renormalizable operators in eq.~\eqref{eq:effectiveW} from renormalizable couplings to messengers $A_1$, $A_2$, $B_1$ and $B_2$ via diagrams shown in fig.~\ref{fig:messengerDiagrams}. The superpotential for the theory including the heavy messengers is
\begin{align}
 W \, &= \, g_1 H^2 A_1 + m_A A_1 A_2 + S \left( g_2 A_2^2 - \Lambda^2 \right) + g_H \Phi H B_1 + g_N \Phi N B_2 + m_B B_1 B_2 + ..., \label{eq:W}
\end{align}
where the dots again denote terms which are irrelevant for inflation. A possible choice of symmetries and charge assignments for all fields including the messengers is given in table~\ref{tab:symmetryChargesExtended}.

The K\"{a}hler potential is again expanded in powers of fields over Planck scale, see eq.~\eqref{eq:K}, with $Y = \left\{ S, H, N, \Phi, A_1,A_2,B_1,B_2 \right\}$.

We will now in turn calculate the inflaton potential $V_{\rm inf}$, the critical inflaton field value $\phi_{\rm c}$ and the vacuum expectation value $\braket{H^2}$ after inflation to compare this models' predictions to the tribrid model of section~\ref{sec:tribrid}.

\begin{table}[bt]
\centering
\begin{tabular}{ | c | C{0.92cm} C{0.92cm} C{0.92cm} C{0.92cm} C{0.92cm} C{0.92cm} C{0.92cm} C{0.92cm} | }
  \hline
  & $S$ & $H$ & $N$ & $\Phi$ & $A_1$ & $A_2$ & $B_1$ & $B_2$ \\
  \hline
  $U(1)_R$ & $2$ & $0$ & $1$ & $1/2$ & $2$ & $0$ & $3/2$ & $1/2$ \\
  $\mathbf{Z}_{4n}$ & $0$ & $n$ & $n-2$ & $n+1$ & $2n$ & $2n$ & $2n-1$ & $2n+1$ \\
  \hline
\end{tabular}
\caption{One possible set of symmetries and charge assignments for the superpotential in eq.~\eqref{eq:W}, for any integer $n \geq 3$.}
\label{tab:symmetryChargesExtended}
\end{table}

\subsection{Inflaton potential $V_{\rm inf}$}
The inflaton potential can be calculated from the supergravity F-term potential\footnote{Inflation is assumed to happen along a D-flat direction, so we can neglect the D-term contributions. Corrections due to deviations from D-flatness are generally negligible, see appendix B of \cite{Antusch:2013toa}. The full supergravity Lagrangian can be found e.g.\ in \cite{Chung:2003fi}.}
\begin{align}
 V_F \, = \, e^{K/\mpl^2}\left( D_i K^{i \overline{j}} D_j^\dagger  - \frac{3}{\mpl^2} \lvert W \rvert^2  \right),\label{eq:VF}
\end{align}
where $ K^{i \overline{j}}$ is the matrix inverse of the K\"{a}hler metric $K_{ \overline{i}j }$, and
\begin{align}
 K_{ \overline{i}j } = \frac{ \partial^2 K }{ \partial Y^\dagger_i \partial Y_j },\quad
 D_i = \frac{ \partial W }{ \partial Y_i } + \frac{W}{\mpl^2} \frac{ \partial K }{ \partial Y_i }. \label{eq:VFcomponents}
\end{align}

During inflation, all non-inflaton fields are stabilized at zero, so the inflaton potential is given by the scalar potential from eq.~\eqref{eq:VF} with $S=H=N=A_i=B_i=W=0$. The expression becomes quite simple due to $D_i = 0$ for all $i\neq S$, and $D_S = W_S = -\Lambda^2$:
\begin{align}
 V_{\rm inf} \, = \, e^{K/\mpl^2} \left( W_S K^{S \overline{S}} W_S^\dagger \right) \, = \, \Lambda^4 e^{K/\mpl^2} K^{S \overline{S}}.
\end{align}
The inverse K\"{a}hler metric can be expanded as a Neumann series:
\begin{align}
\left( K^{i\overline{j}} \right) &= \left( K_{\overline{i}j} \right)^{-1} = \left( \mathbf{1}_8 - \left[ \mathbf{1}_8 - \left( K_{\overline{i}j} \right) \right] \right)^{-1} 
  =  \sum\limits_{k=0}^{\infty} \left[ \mathbf{1}_8 - \left( K_{\overline{i}j} \right) \right]^k. \label{eq:NeumannSeries}
\end{align}
In this expression, $\left( K^{i\overline{j}} \right)$ and $\left( K_{\overline{i}j} \right)$ are $8 \times 8$ matrices, and the multiplication is meant as a matrix multiplication. Only terms proportional to $K_{\overline{S}S}$ contribute, as all other terms in this matrix multiplication (like $K_{\overline{S}Y_i} K_{\overline{Y}_iS}$) are proportional to $S \overline{S} = 0$, so the inverse of the K\"{a}hler metric has the very simple form
\begin{align}
K^{S\overline{S}} &= \sum\limits_{k=0}^{\infty} \left( 1 - K_{\overline{S}S} \right)^k = \frac{1}{K_{\overline{S}S}}.
\end{align}
The inflaton potential is therefore
\begin{align}
 V_{\rm inf} \, &= \, \Lambda^4 \left[ \frac{ e^{K/\mpl^2} }{ K_{\overline{S}S} } \right]_{S=H=...=0} \, \simeq \, \Lambda^4 \left(  1 \, + \, \frac{a}{\mpl^2} \phi^2 \, + \, \frac{b}{\mpl^4} \phi^4 \, + \, ...  \right). \label{eq:kahlerSimpleExtended}
 \end{align}
This tree-level inflaton potential is identical to the result in eq.~\eqref{eq:kahlerSimple1} for the non-renormalizable superpotential. The underlying reason is that during inflation, both models have $W=0$, $\lvert W_S \rvert = \Lambda^2$, and $W_i=0$ for all $i \neq S$. No matter how the superpotential is changed, as long as it contains the term $W \supset -\Lambda^2 S$ and all other terms have at least two powers of non-inflaton fields (which are stabilized at zero during inflation), the inflaton potential will always reduce to eq.~\eqref{eq:kahlerSimpleExtended} during inflation.

\subsection{Critical inflaton field value $\phi_{\rm c}$}

We now want to find the critical inflaton field value $\phi_{\rm c}$, which is defined as the inflaton field value below which the waterfall field develops a tachyonic mass:
\begin{align}
 m_H^2(\phi_c) = 0. \label{eq:defPhiCrit}
\end{align}
The first step will be to find the waterfall field's mass matrix and the second step will be to solve eq.~\eqref{eq:defPhiCrit} for $\phi_c$.

\subsubsection*{Calculation of the waterfall fields' mass terms}
To determine the waterfall fields' mass terms, we need to keep terms in the scalar potential given by eq.~\eqref{eq:VF} up to quadratic order in $H$. We will also need to keep terms up to quadratic order in $B_2$, because we will find some mixing between $H$ and $B_2$. We also keep terms up to leading order in $\Phi$ to find the correct dependence of the mass terms on the inflaton field value.

The renormalizable contributions to the mass matrix can be calculated from $V_F^{\rm (ren)} = \left| W_i \right|^2$. The relevant terms are:
\begin{subequations}
 \begin{align}
  \left| W_N \right|^2 \, &= \, \left| g_N \Phi B_2 \right|^2, \\
  \left| W_{B_1} \right|^2 \, &= \, \left| g_H \Phi H + m_B B_2 \right|^2
  \, = \, \lvert m_B B_2 \rvert^2 + \lvert g_H \Phi \rvert^2 \lvert H \rvert^2 + 2 \operatorname{Re}\left( g_H m_B^\dagger \Phi H B_2^\dagger  \right). \label{eq:WB1quad}
 \end{align}
\end{subequations}
The non-renormalizable corrections to these masses can be calculated starting with the full formula from eq.~\eqref{eq:VF}, expanding $K$ as in eq.~\eqref{eq:K} and keeping only terms up to quadratic order in $Y_i/\mpl$. Their main effect is generating additional Hubble-sized diagonal mass terms for all fields during inflation (see appendix~\ref{appendix:sugraCorrections} for details):
\begin{align}
 V_{\rm SUGRA} \, \simeq \, \frac{c_i \Lambda^4}{\mpl^2} \lvert Y_i \rvert^2 \, \equiv \, \Delta m_i^2 \lvert Y_i \rvert^2 , \label{eq:VsugraIntro}
\end{align}
where the $c_i$ are functions of the K\"{a}hler potential coupling constants $\kappa_{ij}$.

\subsubsection*{Waterfall fields' mass matrix}
Due to the mixing term in eq.~\eqref{eq:WB1quad}, we must consider the entire $H$-$B_2$ mass matrix to determine for which $\phi$ the lowest mass eigenvalue becomes tachyonic.

To make the calculation less cumbersome, we choose $g_H$, $g_N$, $m_B$ and $\Phi$ to be real; their phases can be absorbed into $H$, $N$, $B_1$ and $B_2$ by field redefinitions. We then decompose the fields into real and imaginary parts:
\begin{align}
 B_2 \, = \, \frac{1}{\sqrt{2}}\left(  b_R + i b_I  \right), \quad H \, = \, \frac{1}{\sqrt{2}}\left(  h_R + i h_I  \right), \quad \Phi \, = \, \frac{\phi}{\sqrt{2}}.
\end{align}
The potential for $H$ and $B_2$ is now
\begin{align}
 V_{HB_2} \, &= \, \frac{1}{2} \left( m_B^2 + \Delta m_{B_2}^2 + \frac{ g_N^2 \phi^2 }{ 2 } \right) \left( b_R^2 + b_I^2 \right) + \frac{1}{2} \left( \Delta m_H^2 + \frac{ g_H^2 \phi^2 }{ 2 } \right) \left( h_R^2 + h_I^2 \right)     \notag \\
 &\quad + \frac{g_H \phi}{\sqrt{2}}m_B \left(  b_R h_R + b_I h_I  \right).
\end{align}
The mass matrices for the field pairs $(b_R, h_R)$ and $(b_I, h_I)$ are
\begin{align}
 m^2_{HB_2} \, = \, \left( \begin{array}{cc}
m_B^2 + \Delta m_{B_2}^2 + \frac{ g_N^2 \phi^2 }{ 2 } & \frac{g_H \phi}{\sqrt{2}}m_B \\
\frac{g_H \phi}{\sqrt{2}}m_B & \Delta m_H^2 + \frac{ g_H^2 \phi^2 }{ 2 } \end{array} \right). \label{eq:waterfallMassMatrix}
\end{align}
For large inflaton values, this mass matrix is dominated by the diagonal entries proportional to $\phi^2$:
\begin{align}
  m^2_{HB_2} \, \xrightarrow{ \hphantom{.}\phi \,\rightarrow\, \infty\hphantom{.} } \, \left( \begin{array}{cc}
\frac{ g_N^2 \phi^2 }{ 2 } & 0 \\
0 & \frac{ g_H^2 \phi^2 }{ 2 } \end{array} \right) + ..., \label{eq:HB2stable}
\end{align}
so that all mass eigenvalues are positive for very large inflaton values.\footnote{One can also check that $N_* \sim 50$ e-folds before the end of inflation, when CMB scales cross the horizon, the $H$-$B_2$ mass eigenvalues are large compared to the Hubble scale, so that the waterfall field's perturbations can be neglected for computing the spectrum of primordial curvature perturbations.}

\subsubsection*{Solving for $\phi_c$}

For small inflaton values below $\phi_c$, one of the mass eigenvalues becomes negative. We can determine $\phi_c$ from the condition $m_{HB_2,\pm}^2(\phi_c) = 0$, where $m_{HB_2,\pm}^2$ are the eigenvalues of the waterfall fields' mass matrix in eq.~\eqref{eq:waterfallMassMatrix}. To find a zero eigenvalue, we just have to set the determinant of the matrix to zero:
\begin{align}
 0 \, &= \, \operatorname{det}\left( \begin{array}{cc}
m_B^2 + \Delta m_{B_2}^2 + \frac{ g_N^2 \phi_c^2 }{ 2 } & \frac{g_H \phi_c}{\sqrt{2}}m_B \\
\frac{g_H \phi_c}{\sqrt{2}}m_B & \Delta m_H^2 + \frac{ g_H^2 \phi_c^2 }{ 2 } \end{array} \right) \notag\\
&= \, \left( \frac{g_N^2 \phi_c^2}{2} + m_B^2 + \Delta m_{B_2}^2  \right)\left( \frac{g_H^2 \phi_c^2}{2} + \Delta m_{H}^2  \right) - \frac{g_H^2 \phi_c^2}{2} m_B^2.
\end{align}
We can easily solve this for $\phi_c^2$. The exact result looks somewhat messy, but we find a simple leading-order result by expanding in powers of $\Delta m_i/m_B \sim \Hubble/m_B$:
\begin{align}
 \phi_c^2 \, \simeq \, \frac{2m_B}{g_N g_H} \lvert \Delta m_H \rvert + ...\label{eq:phiCrit}
\end{align}
Up to corrections suppressed by powers of $\Hubble/m_B$, which we denoted by ``$...$'', eq.~\eqref{eq:phiCrit} coincides with eq.~\eqref{eq:kahlerSimple2} from the non-renormalizable theory if we identify
\begin{align}
 \Lambda_\phi \, = \, \frac{m_B}{g_N g_H},
\end{align}
just as one would expect from integrating out the heavy $B_i$ fields in the renormalizable theory.

\subsection{Vacuum expectation values after inflation}
The SUSY-preserving global minimum can be determined by setting all $W_i$ to zero. A straightforward calculation shows that this requires
\begin{subequations}
 \begin{align}
  \braket{A_2^2} \, &= \, \frac{\Lambda^2}{g_2}, \label{eq:globalMinimumA2} \\
  \braket{H^2} \, &= \, -\frac{m_A}{g_1}\braket{A_2} \, = \, \pm \frac{m_A \Lambda}{g_1\sqrt{g_2}}. \label{eq:globalMinimumH}
  \end{align}
\end{subequations}

Eq.~\eqref{eq:globalMinimumH} coincides with eq.~\eqref{eq:kahlerSimple3} from the non-renormalizable tribrid model if
\begin{align}
 \Lambda_H \, = \, \frac{m_A}{g_1 \sqrt{g_2}},
\end{align}
just as one would expect from integrating out the heavy $A_i$ fields in the renormalizable theory.

Most of the other fields are automatically stabilized at zero after inflation. Only two directions do not gain a mass from the operators explicitly shown in eq.~\eqref{eq:W}, but they can easily be made massive by extra operators which are irrelevant during inflation, e.g.\ $\Delta W \propto A_2 N^2$ and  $\Delta W \propto D B_2 A_2$ with an extra chiral superfield $D$.\footnote{As a general rule, such superpotential terms should contain at least three non-inflaton fields and no power of $S$ (which makes them negligible during inflation), and one should check that they do not generate unwanted operators after inflation when $H$ and $A_2$ acquire vacuum expectation values.}

\subsection{Stabilization of messenger fields during inflation}
We have shown that the renormalizable model reproduces the non-renormalizable model as characterized by eqs.~\eqref{eq:kahlerSimple1}--\eqref{eq:kahlerSimple3}, assuming that $S=N=B_1=A_1=A_2=0$ during inflation. We should now check that this assumption is consistent. Note that we have already shown that $H=B_2=0$ for $\phi > \phi_c$, see eq.~\eqref{eq:HB2stable}.

For $N$ and $B_1$, we find mixing just as for $H$ and $B_2$:
\begin{subequations}
 \begin{align}
  \left| W_H \right|^2 \, &= \, \left| g_H \Phi B_1 + 2g_1 H A_1 \right|^2 \, = \, \lvert g_H \Phi B_1 \rvert^2 + ..., \\
  \left| W_{B_2} \right|^2 \, &= \, \left| g_N \Phi N + m_B B_1 \right|^2
  \, = \, \lvert m_B B_1 \rvert^2 + \lvert g_N \Phi \rvert^2 \lvert N \rvert^2 + 2 \operatorname{Re}\left( g_N m_B^\dagger \Phi N B_1^\dagger  \right). \label{eq:WB2quad}
 \end{align}
\end{subequations}
The calculation of the eigenvalues works analogously to the discussion for the $H$-$B_2$ mass matrix. For large $\phi$, the mass matrix is nearly diagonal with very large masses $m_{NB_1}^2 \propto \phi^2$. For smaller $\phi$, a positive supergravity mass $\Delta m_N^2 \gtrsim \Hubble^2$ is sufficient to stabilize all $N$-$B_1$ directions even for $\phi \leq \phi_c$.

The mass terms for $S$, $A_1$ and $A_2$ also take the form $\sum_i ( \lvert W_i \rvert^2 + \Delta m_i^2 )$ during inflation:
\begin{subequations}
\begin{align}
 m_{S}^2 \, &= \, \Delta m_{S}^2, \\
 m_{A_1}^2 \, &= \, m_A^2 + \Delta m_{A_1}^2, \\
 m_{A_2}^2 \, &= \, m_A^2 \pm 2 g_2 \Lambda^2 + \Delta m_{A_2}^2, \label{eq:mA2}
\end{align}
\end{subequations}
where the $\pm$ sign for $A_2$ is ``$-$'' for the real and ``$+$'' for the imaginary component. We see that $A_2=0$ is stable if
\begin{align}
 m_A \, > \, \sqrt{2g_2} \Lambda,  \label{eq:mininumMA}
\end{align}
or if $g_2 \lesssim \mathcal{O}(\Lambda^2/\mpl^2)$ so that $2g_2 \Lambda^2 < \Delta m_{A_2}^2 \sim \mathcal{O}(\Lambda^4/\mpl^2)$.

$A_1$ is strictly heavier than the real component of $A_2$, and $S$ can always have a super-Hubble mass depending on the K\"{a}hler potential (as usual in tribrid inflation).

In summary, we find that if eq.~\eqref{eq:mininumMA} is satisfied or if $g_2$ is very small, all non-inflaton fields can be stabilized at zero during inflation.

\section{One-loop corrections to the inflaton potential}
\label{sec:loop}

In section~\ref{sec:renormalizable}, we have shown that the renormalizable superpotential given by eq.~\eqref{eq:W} including heavy messenger fields $A_i$ and $B_i$ leads to the same tree-level predictions for inflation as the non-renormalizable tribrid superpotential as defined by eq.~\eqref{eq:effectiveW}, up to corrections of order $\Hubble/m_A$, $\Hubble/m_B$. In particular, we demonstrated that both models reduce to hybrid inflation with the same inflaton potential given by eq.~\eqref{eq:kahlerSimple1}, ending with a waterfall at the same critical inflaton field value given by eq.~\eqref{eq:kahlerSimple2}, after which the waterfall field acquires a vacuum expectation value given by eq.~\eqref{eq:kahlerSimple3}.

In this section, we want to compare the one-loop corrections to the effective inflaton potential for the non-renormalizable and the renormalizable superpotential. In particular, we want to show that for small inflaton field values $\phi^2 \ll \Lambda_\phi m_B$, the loop corrections are practically identical, up to small shifts that can be absorbed in the tree-level \kahler potential couplings $\kappa_{\Phi S}$ and $\kappa_{\Phi \Phi}$ or $\kappa_{S\Phi \Phi}$, whereas for larger field values the loop corrections can take on a different functional form. However, for \kahlerdriven tribrid inflation, they turn out to be subdominant in both cases, in which case our tree-level result remains valid and the inflationary dynamics are identical for both cases even for large $\phi^2 \gtrsim \Lambda_\phi m_B$.

\subsection{One-loop potential in tribrid inflation}

The loop effects can be studied using the one-loop effective potential:
\begin{align}
 \Delta V_{\rm loop} \, &= \, \frac{1}{64\pi^2} \sum_i (-1)^{2s_i} \,m_i^4(\phi) \left[  \ln\left( \frac{m_i^2(\phi)}{Q^2} \right) - \frac{3}{2} \right],
 \label{eq:Vloop}
\end{align}
where $m_i(\phi)$ is the mass and $s_i$ the spin of the $i$-th particle degree of freedom, and $Q$ is the $\overline{MS}$ renormalization scale.

In tribrid inflation, the one-loop contribution to the inflaton potential is suppressed by two mechanisms:
\begin{enumerate}
 \item Fermions contribute with a minus sign, so in unbroken SUSY, where one has equal numbers of bosonic and fermionic degrees of freedom with identical masses, $\Delta V_{\rm loop} = 0$. During inflation, SUSY is broken by $\left| W_S \right|^2 \simeq \Lambda^4$, which leads to some mass splittings between scalars and fermions. The resulting loop potential is nevertheless strongly suppressed by partial cancellations between bosonic and fermionic contributions.
 \item Some fields have no $\phi$-dependent mass terms from the superpotential, but only get some Planck-suppressed $\phi$-dependent mass terms from the \kahler potential. In this case, the inflaton-dependent part of the one-loop potential must also be Planck-suppressed.
\end{enumerate}
For $S$ and $A_1$, both suppression mechanisms work simultaneously, and their contribution to the inflaton potential is negligible. The contributions from the inflaton-dependent inflaton mass can also generally be ignored, because the inflaton mass must be small throughout slow-roll: $m_\phi^2 \ll \Hubble^2 \simeq \Lambda^4/(3\mpl^2)$, and therefore $\Delta V_{\rm loop} \ll \Lambda^8/\mpl^4$.

For the non-renormalizable superpotential given by eq.~\eqref{eq:effectiveW}, $H$ and $N$ have strongly $\phi$-dependent masses, so the second suppression mechanism does not apply, and the one-loop potential is
\begin{align}
 \Delta V_{\rm loop}^{(EFT)} \, &\simeq \, \frac{ (\Delta m_H^2 + \Delta m_N^2) }{64\pi^2} \frac{\phi^4}{\Lambda_\phi^2}  \ln\left( \frac{\phi^4}{4 e Q^2 \Lambda_\phi^2} \right). \label{eq:VloopEffective}
\end{align}
For a large part of parameter space, these loop corrections are subdominant to the tree-level inflaton potential of eq.~\eqref{eq:kahlerSimple1} (see \cite{Antusch:2012jc}).

We now want to calculate the one-loop corrections for the renormalizable superpotential of eq.~\eqref{eq:W}. We discuss the corrections due to the $A_2$ masses first, and those due to the $H$-$B_2$ and $N$-$B_1$ mass eigenvalues afterwards. We do not discuss the contributions from $S$, $A_1$ and $\Phi$, which are negligible for the reasons mentioned above.

\subsection{Weakly inflaton dependent mass: $A_2$}
\label{sec:VloopA2}

For $A_2$, the mass splitting between scalar and fermionic components is large, which makes the first suppression mechanism less effective. The scalar and fermionic masses are approximately:
\begin{subequations}
\begin{align}
 m^{(S)\,2}_{A_2}(\phi) \, &\simeq \, \mathcal{N}_A(\phi) \left( m_A^2 \pm 2g_2 \Lambda^2 + \frac{c \Lambda^4}{\mpl^2} \right) \label{eq:massA2s},\\
 m^{(F)\,2}_{A_2}(\phi) \, &\simeq \, \mathcal{N}_A(\phi) \, m_A^2,\label{eq:massA2f}
\end{align}
\end{subequations}
where the $\pm$ sign is ``$-$'' for the real and ``$+$'' for the imaginary scalar component, and
\begin{align}
\mathcal{N}_A(\phi) \, = \, \frac{e^{K/\mpl^2}}{ K_{\overline{A}_1A_1}K_{\overline{A}_2A_2}} \, = \, 1 + \mathcal{O}(\phi^2/\mpl^2)
\end{align}
is a multiplicative mass rescaling from non-renormalizable \kahler potential contributions (see appendix~\ref{appendix:sugraCorrections} for details).

To understand the qualitative behaviour of $\Delta V_{\rm loop}^{(A_2)}$ for these masses, we observe that
\begin{enumerate}
 \item $\Delta V_{\rm loop}^{(A_2)} = 0$ for $\Lambda = 0$.
 \item $\Delta V_{\rm loop}^{(A_2)}$ is an even function of $\Lambda^2$: $\Delta V_{\rm loop}^{(A_2)}(\Lambda^2) = \Delta V_{\rm loop}^{(A_2)}(-\Lambda^2)$.
\end{enumerate}
This implies that $\Delta V_{\rm loop}^{(A_2)}$ can be expanded in powers of $\Lambda^4$, with no term containing less than one power of $\Lambda^4$.

The only inflaton dependence arises from expanding $\mathcal{N}(\phi)$ in powers of $\phi^2/\mpl^2$, so the inflaton-dependent part of $\Delta V_{\rm loop}$ generally has the form\footnote{Higher powers of $\Lambda^4$, like $\Lambda^8/m_A^4$, can be absorbed in the coefficients $\tilde{a}$ and $\tilde{b}$.}
\begin{align}
 \Delta V_{\rm loop}^{(A_2)} \, = \, \Lambda^4 \left( \tilde{a} \frac{\phi^2}{\mpl^2} + \tilde{b} \frac{\phi^4}{\mpl^4} + ... \right),\label{eq:VloopA2}
\end{align}
which has the same form as the tree-level expansion in eq.~\eqref{eq:kahlerSimpleExtended}. The loop corrections can therefore be absorbed in the tree-level couplings, replacing $a \rightarrow a_{\rm eff} = (a + \tilde{a})$ and $b \rightarrow b_{\rm eff} = (b + \tilde{b})$, which is equivalent to small shifts in the \kahler potential couplings $\kappa_{\Phi S}$ and $\kappa_{\Phi\Phi}$ or $\kappa_{\Phi\Phi S}$ in the tree-level calculation.

The numeric factors $\tilde{a}$ and $\tilde{b}$ are also suppressed by the loop factor $\log(...)/64\pi^2$. For the one-loop approximation to be valid, this loop factor must be small (otherwise two and more loops would be expected to give even larger corrections), and one generally finds $\tilde{a}, \tilde{b} \ll 1$. However, even small corrections $\tilde{a} \sim 10^{-2}$ lead to measurable changes in the predictions, as the predictions of tribrid inflation are very sensitive to the precise value of $a$.

\subsection{Strongly inflaton dependent mass: $H$, $N$ and $B_i$}

We now want to discuss the $H$-$B_2$ and $N$-$B_1$ directions whose masses are strongly $\phi$-dependent due to their renormalizable couplings involving the inflaton field. We can focus on the $H$-$B_2$ direction with the mass matrix given in eq.~\eqref{eq:waterfallMassMatrix}; the calculation for $N$-$B_1$ is identical except for a substitution $g_N \leftrightarrow g_H$, $\Delta m_H^2 \leftrightarrow \Delta m_N^2$ and $\Delta m_{B_1}^2 \leftrightarrow \Delta m_{B_2}^2$.

The masses take a form similar to eqs.~\eqref{eq:massA2s}--\eqref{eq:massA2f}, but without the large splitting between the real and imaginary scalar components. We write
\begin{subequations}
\begin{align}
 m_{i}^{(F)\,2}(\phi) \, &= \, M_i^2(\phi),\\
 m_{i}^{(S)\,2}(\phi) \, &= \, M_i^2(\phi) \, + \, \delta_i^2,
\end{align}
\end{subequations}
where $\delta_i^2 \sim \mathcal{O}(\Lambda^4/\mpl^2)$ is the SUSY breaking mass splitting term. During inflation, the masses are large compared to the Hubble scale\footnote{Very close to the critical point, $\phi \simeq \phi_c$, two of the masses become small just before the tachyonic instability develops at $\phi = \phi_c$, but during most of inflation the masses are very large. In particular, at the time when CMB scales leave the horizon, $m_{HB_2} \gg \Hubble$ is a very good approximation. This is important because predictions for the primordial spectrum depend most strongly on the potential around the time of horizon crossing, so that is the time when we want our calculation of the loop corrections to be most accurate.}, so we can expand $V_{\rm loop}^{(i)}$ (the contribution of the $i$-th superfield to $V_{\rm loop}$) in powers of $\delta_i^2/M_i^2$:
\begin{align}
 \Delta V_{\rm loop}^{(i)} \, &= \, \frac{1}{64\pi^2} \sum_{\rm s, \, f} (-1)^{2s_i} \,m_i^4(\phi) \left[  \ln\left( \frac{m_i^2(\phi)}{Q^2} \right) - \frac{3}{2} \right] \notag \\
 &= \, \frac{1}{32\pi^2} \left\{ \left( M_i^2 + \delta_i^2 \right)^2 \ln\left( \frac{M_i^2 + \delta_i^2}{e^{3/2}Q^2} \right) - M_i^4 \ln\left( \frac{M_i^2}{e^{3/2}Q^2} \right)\right\} \notag\\
 &= \, \frac{ M_i^2 \delta_i^2 }{16\pi^2}  \ln \left( \frac{M_i^2}{e Q^2} \right) \, + \, \mathcal{O}(\delta_i^4). \label{eq:VloopSimplified}
\end{align}
where the sum in the first line goes over both scalar and fermionic degrees of freedom corresponding to the $i$-th superfield.

We now want to evaluate eq.~\eqref{eq:VloopSimplified} for the $H$-$B_2$ mass matrix. As the $M_i^2$ and $\delta_i^2$ for the mass eigenvalues look quite complicated, we will consider the limits of small and large inflaton values, expanding $M_i^2$ and $\delta_i^2$ in powers of $\phi^2/m_B^2$ for small $\phi$ and in powers of $m_B^2/\phi^2$ for large $\phi$. For $g_N \sim g_H$, these two cases correspond to $\phi^2 \ll \Lambda_\phi m_B$ and $\phi^2 \gg \Lambda_\phi m_B$.

\subsubsection*{Small field values: $\phi^2 \ll \Lambda_\phi m_B$}
For $\phi \ll m_B/g_N$ and $\phi \ll m_B/g_H$, the mass eigenvalues can be expanded in powers of $\phi$. The leading order terms are
\begin{subequations}
\begin{align}
 M_{1}^2 \, &\simeq \, \frac{\phi^4}{4\Lambda_\phi^2}, &   \delta_{1}^2 \, &\simeq \, \Delta m_{H}^2,\\ M_{2}^2 \, &\simeq \, m_B^2 + \frac{g_H^2 + g_N^2}{2} \phi^2, &  \delta_{2}^2 \, &\simeq \, \Delta m_{B_2}^2 + \frac{g_H^2 \phi^2}{2m_B^2}\left( \Delta m_H^2 - \Delta m_{B_2}^2 \right),
\end{align}
\end{subequations}
and eq.~\eqref{eq:VloopSimplified} becomes, to leading order in $\phi^2/m_B^2$:
\begin{subequations}
\begin{align}
 \Delta V_{\rm loop}^{(1)} \, &\simeq \, \frac{\Delta m_H^2}{64\pi^2} \frac{\phi^4}{\Lambda_\phi^2}  \ln\left( \frac{\phi^4}{4e Q^2 \Lambda_\phi^2} \right), \label{eq:VloopH}\\
  \Delta V_{\rm loop}^{(2)} \, &\simeq \, \frac{\phi^2}{32\pi^2} \left[ g_H^2\left( \Delta m_{B_2}^2 - \Delta m_H^2 \right) + \left( g_N^2 \Delta m_{B_2}^2 + g_H^2 \Delta m_H^2 \right) \ln\left( \frac{m_B^2}{Q^2} \right) \right] .\label{eq:VloopB}
\end{align}
\end{subequations}

The smaller mass eigenvalue in this limit is identical to the mass of $H$ for the non-renormalizable superpotential from eq.~\eqref{eq:effectiveW}, up to corrections suppressed by higher powers of $\phi^2/(m_B \Lambda_\phi)$. For this reason, eq.~\eqref{eq:VloopH} reproduces the loop potential for the non-renormalizable tribrid superpotential as given in eq.~\eqref{eq:VloopEffective}.\footnote{We also reproduce the term proportional to $\Delta m_N^2$ when we consider the contribution of the $N$-$B_1$ mass eigenvalues.}

The corrections due to the heavy $B_2$ field, given by eq.~\eqref{eq:VloopB}, can be made small by choosing a suitable renormalization scale $Q \sim m_B$. In principle, $Q$ can be chosen such that the entire bracket is zero (in that case, the leading term is $\mathcal{O}(\phi^4/m_B^4)$). Even if it provides a contribution to the effective inflaton potential, it takes the same functional form as eq.~\eqref{eq:VloopA2}, and could therefore be absorbed in the inflaton potential terms $a_{\rm eff}$ and $b_{\rm eff}$ as discussed for $\Delta V_{\rm loop}^{(A_2)}$ in section~\ref{sec:VloopA2}.

\subsubsection*{Large field values: $\phi^2 \gg \Lambda_\phi m_B$}
For very large $\phi$, the mass eigenvalues asymptote to eq.~\eqref{eq:HB2stable}, with $\delta^2$ given by $\Delta m_H^2$ or $\Delta m_{B_2}^2$. This leads to a loop potential of the form
\begin{align}
 \Delta V_{\rm loop}^{(1,2)} \, = \, \frac{g_{H,N}^2 \Delta m_{i}^2}{32\pi^2} \phi^2 \ln\left( \frac{ g_{N,H}^2 \phi^2 }{ 2eQ^2 } \right),
\end{align}
with a functional form $\Delta V \propto \phi^2 \ln(\phi)$ which is qualitatively different from that of the EFT loop potential in eq.~\eqref{eq:VloopEffective}, and which cannot be absorbed in the tree-level couplings $a$ and $b$ due to the logarithmic dependence. Note that this loop potential is suppressed by $(m_B^2/\phi^2) \ll 1$ compared to the EFT loop potential in eq.~\eqref{eq:VloopEffective}, so it is negligible in cases where even the EFT loop corrections are subdominant.

\subsection{Conclusion on one-loop corrections}
The above calculations indicate that for small inflaton field values $\phi^2 \ll \Lambda_\phi m_B$, the one-loop corrections are almost identical for the non-renormalizable EFT superpotential and the renormalizable superpotential involving heavy messenger fields. Though the heavy messenger fields can introduce some corrections to the effective inflaton potential, the deviations have the same form as the tree-level potential and can be absorbed in the tree-level couplings as (quite small) shifts in the \kahler potential couplings $\kappa_{\Phi S}$ and $\kappa_{\Phi\Phi}$ or $\kappa_{\Phi\Phi S}$.

For large inflaton field values $\phi^2 \gg \Lambda_\phi m_B$, we found that the logarithmic one-loop corrections have a different functional form for the renormalizable and non-renormalizable superpotentials, with $\Delta V_{\rm loop}$ for the renormalizable superpotential suppressed by $\mathcal{O}(m_B^2/\phi^2) \ll 1$ compared to the EFT. 

In general, this shows that the one-loop corrections for inflaton field values above the cutoff scale can deviate between the renormalizable and the non-renormalizable superpotential, which could be relevant when constructing explicit models. However, in the many cases where even the (larger) EFT one-loop corrections are small compared to the tree-level inflaton potential, our analysis indicates that the inflationary dynamics for both superpotentials are determined by their tree-level predictions and therefore identical up to $\mathcal{O}(\Hubble/m_A)$.

\section{Generalisation to other superpotentials and messenger topologies}
\label{sec:generalisation}

In the previous section, we have discussed how to generate the specific non-renormalizable tribrid superpotential defined in eq.~\eqref{eq:effectiveW} using the renormalizable superpotential in eq.~\eqref{eq:W}, such that both superpotentials lead to the same inflationary dynamics even for large field values $\phi$ above the suppression scales $\Lambda_\phi$ and $\Lambda_H$ of the non-renormalizable superpotential.

In this section, we briefly discuss how this explicit worked-out example can be generalized to more general tribrid superpotentials and how different choices for the messenger sector can affect the inflationary predictions.

\subsection{More general superpotentials}

It is possible to extend our analysis to more general \kahlerdriven tribrid superpotentials of the form
\begin{align}
 W \, = \, S\left( \frac{H^\ell}{\Lambda_H^{\ell-2}} - \Lambda^2 \right) \, + \, \frac{1}{\Lambda_\phi^{n-1}} H N \Phi^n, \label{eq:generalW}
\end{align}
with $\ell \geq 4$ and $n \geq 2$. The last term can also be replaced by $H^2 \Phi^n/\Lambda_\phi^{n-1}$ without changing the inflationary dynamics, except for a numerical factor of $2$ in the waterfall mass. It is also straightforward to replace $\Phi^n$ by a D-flat direction of multiple fields, e.g.\ $\Phi^2 \rightarrow LH_u$ or $\Phi^3 \rightarrow L H_d E$.

\begin{figure}[bt]
  \centering$
\begin{array}{ccc}
\includegraphics[height=0.166\textwidth]{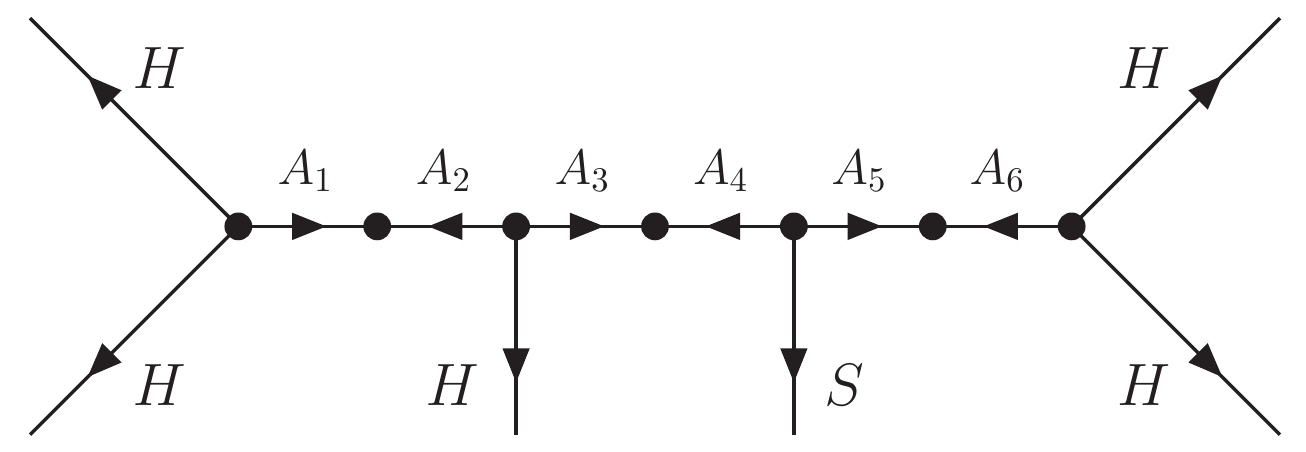} & \hphantom{.....} &
\includegraphics[height=0.166\textwidth]{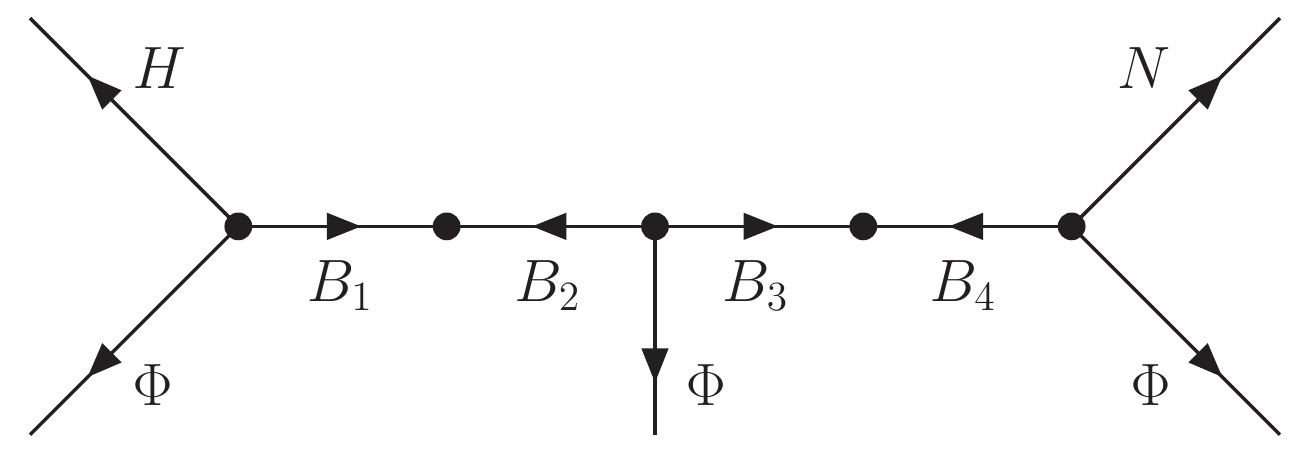}
\end{array}$
  \caption{Diagrams for generating the non-renormalizable operators of the superpotential in eq.~\eqref{eq:generalW} from renormalizable superpotential couplings to heavy messenger fields $A_i$ and $B_i$.}
  \label{fig:messengerDiagrams2}
\end{figure}

For these more general tribrid superpotentials, it is possible to proceed analogously to the discussion in section~\ref{sec:renormalizable} using diagrams like those shown in fig.~\ref{fig:messengerDiagrams2} (with the caveat discussed in section~\ref{sec:additionalOperators}). From the diagrams, one can read off the messengers' symmetry charges, as the total charge of the fields connected to each vertex must be $2$ for an $U(1)_R$ symmetry, $0$ for $U(1)$ symmetries and some multiple of $n$ for $\mathbf{Z}_n$ symmetries.

Higher $\ell > 4$ can generally be achieved by adding more external $H$ legs to the left diagram in fig.~\ref{fig:messengerDiagrams2}. For this kind of messenger topology, the inflationary dynamics is unaffected except for the vacuum expectation value after inflation, which is $\braket{H^\ell} \sim c\Lambda^2$ with a constant $c$ composed of the $A_i$ fields' masses and their couplings to $S$ and $H$. The basic reason is that such diagrams are based only on superpotential couplings of the type $m A_i A_j$ and $H^2 A_i$, which have no effect during inflation, and $S A_i A_j$, which generates a mass splitting for $A_i$ and $A_j$ only, but no mass terms for $H$ which is the field responsible for initiating the waterfall instability.

Larger $n$ can be achieved by adding more external $\Phi$ legs to the right diagram in fig.~\ref{fig:messengerDiagrams2}. This leads to more complicated mixing. In the example given in fig.~\ref{fig:messengerDiagrams2}, we find mixing terms between $H$, $B_2$ and $B_4$, so that the calculation of the mass eigenvalues is more involved, and therefore both the calculation of $\phi_c$ and the estimation of the one-loop corrections are more difficult.

\subsection{Effect of different messenger topologies}

\begin{figure}[bt]
  \centering$
\begin{array}{ccc}
\includegraphics[height=0.166\textwidth]{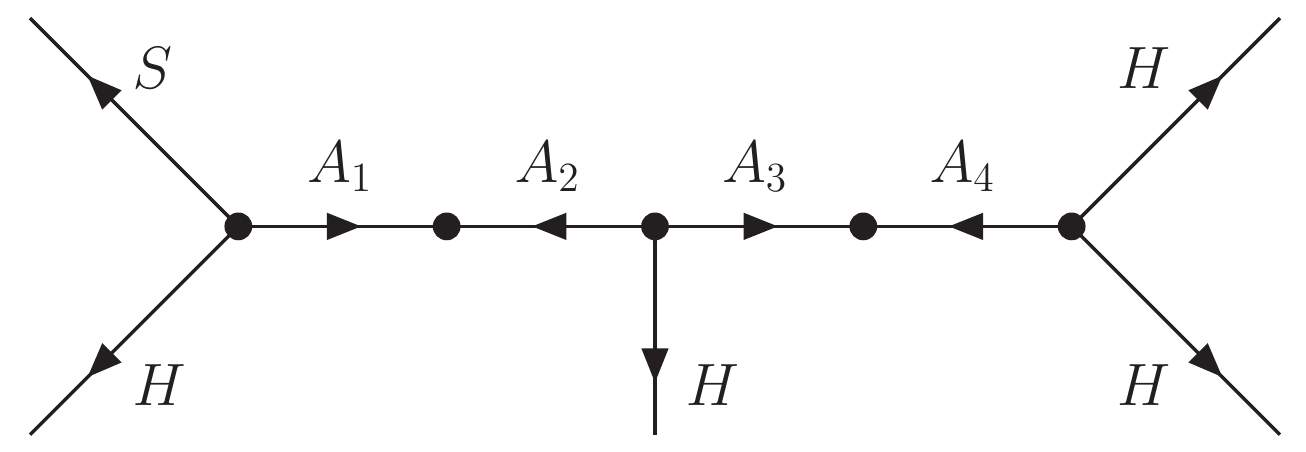} & \hphantom{.....} &
\includegraphics[height=0.166\textwidth]{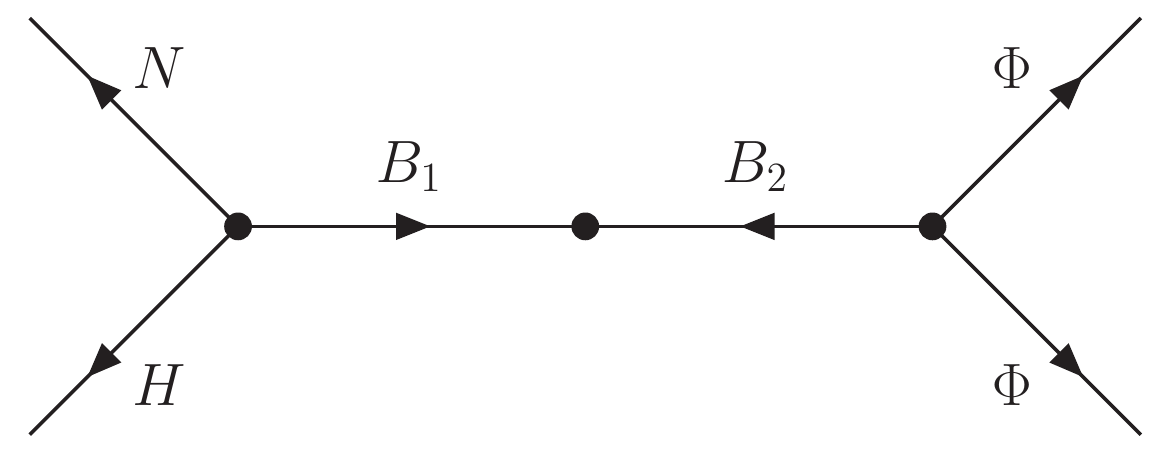}
\end{array}$
  \caption{Alternative diagrams for generating the non-renormalizable operators of the superpotential defined in eq.~\eqref{eq:effectiveW} from different renormalizable couplings to heavy messenger fields $A_i$ and $B_i$. These choices introduce stronger deviations from the predictions of the EFT superpotential in eq.~\eqref{eq:effectiveW}. The left diagram introduces significant $H$-$A_1$ mixing, which can change both $\phi_c$ and $\Delta V_{\rm loop}$. The right diagram changes the dynamics even more drastically, as it destabilizes the $B_1$ field during inflation, which would lead to multi-field inflation involving $\Phi$ and $B_1$.}
  \label{fig:messengerDiagrams3}
\end{figure}

In our analysis, we have focused on a particular choice for the messenger sector which leads to matching predictions between the renormalizable and non-renormalizable superpotential. One could also think of generating the non-renormalizable operators in eq.~\eqref{eq:effectiveW} with different messenger sectors, e.g.\ using the diagrams shown in fig.~\ref{fig:messengerDiagrams3}. However, these alternative messenger sectors introduce stronger deviations from the predictions of the non-renormalizable superpotential defined in eq.~\eqref{eq:effectiveW}.

The left diagram in fig.~\ref{fig:messengerDiagrams3} depicts an alternative way to generate the coupling $S H^4 / \Lambda_H^{2}$. The main difference is that it contains a vertex $SHA_1$, which introduces a large $H$-$A_1$ mixing term in the scalar mass matrix. This mixing affects the mass eigenvalues of the waterfall directions and thus changes both $\phi_c$ and $\Delta V_{\rm loop}$.

The right diagram in fig.~\ref{fig:messengerDiagrams3}, which one might expect to generate $\Delta W \supset HN\Phi^2/\Lambda_\phi$, changes the dynamics even more drastically, as it destabilizes $B_1$ during inflation (the F-term $\lvert W_{B_2}\rvert^2$ is minimized only for $B_1 \sim \Phi^2/m_B$). In general, this will lead to multi-field inflation involving $\Phi$ and $B_1$.\footnote{Even if $B_1$ is heavy during inflation, so that it tracks its minimum at $B_1 \sim \Phi^2/m_B$, the effective inflation potential is changed by canonical normalization of the adiabatic direction along the two-field trajectory, which is a large effect at least for $\Phi \gtrsim \mathcal{O}(m_B)$.}

We do not want to discuss these models in detail, but we want to emphasize that for messenger sectors like those in fig.~\ref{fig:messengerDiagrams3}, the predictions for inflation must be calculated carefully for the model including all the messengers, and the results will generally differ from those expected for the non-renormalizable tribrid superpotentials defined in eq.~\eqref{eq:generalW}.

\subsection{Checking for additional operators allowed by the symmetries}
\label{sec:additionalOperators}

When constructing explicit models including the messenger sector, it is necessary to check that no unwanted additional operators are allowed by the symmetries. This puts some constraints on the possibilities for building messenger sectors. As an example, consider constructing a messenger sector for the superpotential
\begin{align}
 W_{\rm EFT} \, = \, S\left( \frac{H^4}{\Lambda_H^2} - \Lambda^2 \right) \, + \, \frac{1}{\Lambda_\phi} H^2 \Phi^2. \label{eq:problematicW}
\end{align}
This is the same superpotential as in eq.~\eqref{eq:effectiveW}, apart from replacing $HN \rightarrow H^2$, so one might try to construct a messenger sector using the diagrams in fig.~\ref{fig:messengerDiagrams} with $N$ replaced by $H$:
\begin{align}
 W_{\rm ren} \, &= \, g_1 H^2 A_1 + m_A A_1 A_2 + S \left( g_2 A_2^2 - \Lambda^2 \right) + g_H \Phi H B + \frac{m_B}{2} B^2 + ... \label{eq:problematicW2}
\end{align}
However, any choice of $U(1)_R$ and $\mathbf{Z}_n$ symmetries consistent with eq.~\eqref{eq:problematicW2} also allows the troublesome operator
\begin{align}
 \Delta W_{\rm trouble} \, &\propto \, A_2 \Phi^2,
\end{align}
which destabilizes $A_1$ during inflation and generates a tree-level contribution to the inflaton potential.

To be safe, one should always check that the symmetries of the model including the messenger sector do not allow for any additional superpotential operators with less than three powers of non-inflaton, non-$S$ fields. In that case, the extra operators cannot generate any mass terms during inflation, so that no fields can be destabilized and $\phi_c$ is unchanged.

\section{Summary}

In this paper, we have studied whether tribrid inflation can be successfully realized even in the presence of superpotential operators with low cutoff scales $\Lambda_{H}, \Lambda_{\phi} \lesssim \Delta \phi \ll \mpl$, or if inflation must be studied in a more UV complete model explicitly including all particles with masses $m_i \lesssim \Delta \phi$.

We started by constructing a particular UV extension in which the non-renormalizable operators with sub-Planckian suppression scales $\Lambda_{\rm cutoff}$ are replaced by renormalizable couplings to heavy messenger fields. We found that at tree level, the inflationary dynamics for both superpotentials could be reduced to eqs.~\eqref{eq:kahlerSimple1}--\eqref{eq:kahlerSimple3}, even for $\Delta \phi > \Lambda_{H}, \Lambda_{\phi}$. In particular, this implies that the tree-level predictions are identical regardless of whether one uses the non-renormalizable superpotential (with messenger fields already integrated out) or the renormalizable superpotential (explicitly including all messenger fields), apart from small corrections of order $\Hubble/m_A$.

However, the one-loop quantum corrections to the effective inflaton potential are different when calculated with the renormalizable superpotential. Most importantly, the logarithmic corrections due to the waterfall field's mass are suppressed by $(m_B^2/\phi^2)$ for large field values $\phi^2 \gg \Lambda_\phi m_B$ compared to the non-renormalizable superpotential. The heavy messengers' masses also generate small polynomial corrections to the inflaton potential, but those can easily be accounted for by small shifts of the tree-level \kahler potential couplings $\kappa_{\Phi S}$ and $\kappa_{\Phi\Phi}$ or $\kappa_{\Phi\Phi S}$. Our results imply that in the cases in which one-loop corrections are subdominant even for the non-renormalizable superpotential, they will generally be negligible for the renormalizable superpotential as well, so inflation is well-described by the tree-level predictions which are identical for both superpotentials. However, for models in which the one-loop corrections are important, e.g.\ loop-driven tribrid inflation \cite{Antusch:2008pn,Antusch:2009ef,Antusch:2010mv,Antusch:2010va}, it is important to explicitly include all messengers with masses $m_i \lesssim \Delta \phi$.

Finally, we also discussed how our analysis can be extended to more general tribrid superpotentials given by eq.~\eqref{eq:generalW}. We discussed how one can systematically construct a messenger sector to generate the non-renormalizable superpotential operators analogously to our explicit example. We also discussed which qualitative differences occur when using the different messenger topologies given in fig.~\ref{fig:messengerDiagrams3}, and why it is important to check explicitly that the messenger sector does not generate additional unwanted operators which might disturb the inflationary dynamics.

In summary, we have shown that it is possible to realize \kahlerdriven tribrid inflation in particle physics models even when the superpotential contains non-renormalizable operators with suppression scales $\Lambda_{\rm cutoff} \lesssim \Delta \phi$, and we have outlined how a messenger sector can be constructed such that the full theory including the messenger sector leads to the same predictions as the non-renormalizable superpotential. For loop-driven tribrid inflation, our results suggest that the messenger sector must be included explicitly, and that the inflaton potential generally depends on the details of the messenger fields' couplings. These results have important implications for embedding tribrid inflation within realistic particle physics theories that sponsor some intermediate scale $\Lambda_{\rm NP} \ll \mpl$, and they provide useful guidelines for whether it is sufficient to use the simpler effective theory with cutoff scale $\Lambda_{\rm NP}$ or whether it is necessary to construct the messenger sector explicitly to study inflation within any given model.

\appendix
\section*{Appendix}
\section{Supergravity corrections to the mass matrix during inflation}
\label{appendix:sugraCorrections}
In this appendix, we calculate the Planck-suppressed corrections to the mass matrices of scalars and fermions. The purpose of this appendix is to show that during inflation, the corrections take the form given by eq.~\eqref{eq:VsugraIntro} for scalars and that they are negligible for fermions. The calculations in this appendix can be easily generalized to other messenger sectors if the new messenger fields satisfy $W = W_i = 0$ (for $i \neq S$) during inflation and the \kahler potential has the form of eq.~\eqref{eq:K}.

In this section, we set the reduced Planck mass $\mpl=1$ to keep the notation as simple as possible.

\subsection{Supergravity potential for the scalar mass matrix}
First we discuss the supergravity corrections $V_{\rm SUGRA}$ to the scalar mass matrix as defined in eq.~\eqref{eq:VsugraIntro}.

Formally, we can find the relevant terms by a power series expansion in $\xi$, with $A_i$, $B_i$, $H$, $N$, $S = \mathcal{O}(\xi)$, neglecting all terms of $\mathcal{O}(\xi^3)$. We evaluate the individual building blocks defined in eq.~\eqref{eq:VFcomponents} for the superpotential in eq.~\eqref{eq:W} and the K\"{a}hler potential in eq.~\eqref{eq:K}:
\begin{subequations}
 \begin{alignat}{3}
  \lvert W \rvert^2 \, &= \, \Lambda^4\lvert S \rvert^2 \, + \, \mathcal{O}(\xi^3) \, &= \, \mathcal{O}(\xi^2), \\
  W_S \, &= \, -\Lambda^2 + g_2 A_2^2 \, &= \, \mathcal{O}(\xi^0),\\
  W_H \, &= \, 2g_1 A_1 H + g_H \Phi B_1 \, &= \, \mathcal{O}(\xi^1), \\
  W_N \, &= \, g_N \Phi B_2 \, &= \, \mathcal{O}(\xi^1), \\
  W_{A_1} \, &= \, g_1 H^2 + m_A A_2 \, &= \, \mathcal{O}(\xi^1), \\
  W_{A_2} \, &= \, 2g_2 S A_2 + m_A A_1 \, &= \, \mathcal{O}(\xi^1), \\
  W_{B_1} \, &= \, g_H \Phi H + m_B B_2 \, &= \, \mathcal{O}(\xi^1),\\
  W_{B_2} \, &= \, g_N \Phi N + m_B B_1 \, &= \, \mathcal{O}(\xi^1),\\
  W_\Phi \, &= \, g_H H B_1 + g_N N B_2 \, &= \, \mathcal{O}(\xi^2), \\
  W K_\Phi \, &= \, -\Lambda^2 S \Phi^\dagger \left(  1 + 2\kappa_{\Phi\Phi}\lvert\Phi\rvert^2 + ... \right) \, &= \, \mathcal{O}(\xi^1), \\
  W K_{i \neq \Phi} \, &= \, -\Lambda^2 S Y_i^\dagger \left(  1 + \kappa_{\Phi i}\lvert\Phi\rvert^2 + ... \right) + \mathcal{O}(\xi^3) \, &= \, \mathcal{O}(\xi^2).
 \end{alignat}
\end{subequations}
Note that this implies that $D_i = \mathcal{O}(\xi^1)$ for all $i \neq S$, and $D_S = \mathcal{O}(\xi^0)$.

The inverse K\"{a}hler metric $K^{ i\overline{j} }$ is always contracted with $D_i D_j^\dagger$ in eq.~\eqref{eq:VF}. For $i \neq S \neq j$, for which $D_i D_j^\dagger = \mathcal{O}(\xi^2)$, we therefore only need terms up to $\mathcal{O}(\xi^0)$, for $i \neq S = j$ terms up to order $\mathcal{O}(\xi^1)$, and only for the diagonal element $i=j=S$ we need terms up to $\mathcal{O}(\xi^2)$.

With this in mind, we can use eq.~\eqref{eq:NeumannSeries} to expand $K^{ i\overline{j} }$ up to the required order in $\xi$. The diagonal element for $S$ to order $\mathcal{O}(\xi^2)$ can be calculated as a Neumann series (defining $\Delta K_{\overline{i}j} = K_{\overline{i}j} - \delta_{ij}$):
\begin{align}
 K^{S\overline{S}} \, &= \, 1 -\Delta K_{\overline{S}S} + \sum_{\mathclap{i}} \Delta K_{\overline{S}i} \Delta K_{\overline{i}S} - ... \notag\\
 &= \, 1 - \sum_{i} \left( \kappa_{S i} + (\kappa_{S \Phi i}-2\kappa_{Si}\kappa_{S\Phi}) \lvert \Phi \rvert^2 \right)(1 + 3\delta_{Si}) \lvert Y_i \rvert^2 + \kappa_{S\Phi}^2 \lvert \Phi \rvert^2 \lvert S \rvert^2 + ...,
\end{align}
where the $...$ denote terms of $\mathcal{O}(\xi^3)$ and $\mathcal{O}(\Phi^4)$ which we neglected. The diagonal element for $a \neq S$ to order $\mathcal{O}(\xi^0)$ is:
\begin{align}
 K^{a\overline{a}} \, &= \, 1 -\Delta K_{\overline{a}a} + \sum_{\mathclap{i}} \Delta K_{\overline{a}i} \Delta K_{\overline{i}a} - ... \notag\\
 &= \, 1 - \kappa_{\Phi a} (1 + 3\delta_{\Phi a}) \lvert \Phi \rvert^2 + ... + \mathcal{O}(\xi^2).
\end{align}
For the off-diagonal elements $a \neq b$, we have:
\begin{align}
 K^{a\overline{b}} \, &= \, -\Delta K_{\overline{a}b} + \sum_{\mathclap{i}} \Delta K_{\overline{a}i} \Delta K_{\overline{i}b} - ... \notag\\
 &= \, -\left( \kappa_{ab} + \kappa_{\Phi ab}\left( 1 + \delta_{\Phi a} + \delta_{\Phi b} \right) \left| \Phi \right|^2 \right) Y_a Y_b^\dagger + Y_a Y_b^\dagger \sum_{\mathclap{i}} \kappa_{ai} \kappa_{bi} (1 + \delta_{ai} + \delta_{bi}) \lvert Y_i \rvert^2 - ... \notag\\
 &= \, Y_a Y_b^\dagger \left(  -\kappa_{ab} + (\kappa_{\Phi a}\kappa_{\Phi b} - \kappa_{\Phi a b})(1 + \delta_{\Phi a} + \delta_{\Phi b}) \lvert \Phi \rvert^2 + ...  \right) \, + \, \mathcal{O}(\xi^3).
\end{align}
This is of order $\mathcal{O}(\xi^1)$ if $a = \Phi$ or $b = \Phi$, and order $\mathcal{O}(\xi^2)$ otherwise. Due to $D_i = \mathcal{O}(\xi^1)$ for all fields $i \neq S$, none of the off-diagonal elements contribute to the mass matrix except $D_S K^{ S\overline{\Phi} } D_\Phi^\dagger + \text{h.c.}$, which generates a small inflaton-dependent mass for $S$:
\begin{align}
 D_\Phi K^{ \Phi\overline{S} } D_S^\dagger \, &= \, W K_\Phi K^{ \Phi\overline{S} } W_S  + \mathcal{O}(\xi^3) \notag\\
 &= \, -\Lambda^2 S \Phi^\dagger \left(  1 + ... \right)\Phi S^\dagger \left(  -\kappa_{S\Phi} + ...  \right) \left( -\Lambda^2 \right) + \mathcal{O}(\xi^3) \notag\\
 &= -\kappa_{S\Phi} \, \Lambda^4 \lvert \Phi \rvert^2 \lvert S \rvert^2 + ...
\end{align}
We evaluate the diagonal terms in turns for $\Phi$, $S$ and other fields $Y_i$. We start with $\Phi$:
\begin{align}
 D_\Phi K^{\Phi\overline{\Phi}} D_\Phi^\dagger \, &= \, \lvert W K_\Phi \rvert^2 + \mathcal{O}(\xi^3)\notag\\
 &= \, \Lambda^4 \lvert \Phi \rvert^2 \lvert S \rvert^2 + ... \, + \, \mathcal{O}(\xi^3).
\end{align}
For $S$, we have
\begin{align}
 D_S K^{S\overline{S}} D_S^\dagger \, &= \, \lvert D_S \rvert^2 \left[  1 - \sum_{i} \left( \kappa_{S i} + \kappa_{S \Phi i} \lvert \Phi \rvert^2 \right)(1 + \delta_{Si}) \lvert Y_i \rvert^2 +\left( \kappa_{S\Phi}^2 + ... \right) \lvert \Phi \rvert^2 \lvert S \rvert^2  \right] + \mathcal{O}(\xi^3)\notag\\
 &= \, \left| g_2 A_2^2 - \Lambda^2( 1 + S\Phi^\dagger + ... ) \right|^2 \left( 1 + \sum_i (c_i + d_i \lvert\Phi\rvert^2 + ...) \lvert Y_i \rvert^2 + ... \right) + \mathcal{O}(\xi^3) \notag\\
 &= \Lambda^4 + \sum_i \Delta m_i^2 \lvert Y_i \rvert^2 - 2 g_2 \Lambda^2 \operatorname{Re}\left( A_2^2 \right) \, + \, \mathcal{O}(\xi^3),
\end{align}
with supergravity mass terms $\Delta m_i^2 = (c_i + d_i \left| \Phi \right|^2 + ...) \Lambda^4$, where the coefficients $c_i$ and $d_i$ are functions of the $\kappa_{ij}$ and $\kappa_{ijk}$. For the other fields $Y_i$ ($i \neq \Phi$), we recover the simple form
\begin{align}
 D_{i} K^{i\overline{i}} D_i^\dagger \, &= \, \lvert W_i \rvert^2 \left( 1 - \kappa_{\Phi i} \lvert \Phi \rvert^2 + ... \right)  \, + \, \mathcal{O}(\xi^3). \label{eq:sugraRescaling1}
\end{align}
Another two sources of SUGRA corrections are the $\lvert W \rvert^2$ term in eq.~\eqref{eq:VF}, which introduces another contribution to the mass term of $S$ that can be absorbed in the constant $c_S$, and the exponential $e^{K}$, which also contributes to the mass terms via
\begin{align}
 e^K \lvert D_S \rvert^2 = (1 + K + ...)(\Lambda^4 + ...).
\end{align}
Correctly expanding the exponential, one finds additional contributions to the masses of all fields, which are again of order $\Delta m_i^2 \sim \mathcal{O}(\Lambda^4)$. These can also be absorbed in the definition of the $c_i$ and $d_i$ above.

To calculate the precise form of the $d_i$, one must also take into account the effect of canonical normalization, which during inflation can be achieved by a redefinition $Y_i \rightarrow Y_i/\sqrt{K_{\overline{i}i}} = (1 - \frac{\kappa_{i\Phi}}{2} \lvert \frac{\Phi}{\mpl} \rvert^2 + ...)Y_i$ for the non-inflaton scalar fields $Y_i$, which generates inflaton-dependent mass terms from the non-inflaton-dependent mass terms, i.e.\ $d_i \rightarrow d_i -\kappa_{i\Phi} c_i$.\footnote{See \cite{Antusch:2012jc} for details on canonical normalization during tribrid inflation.}

Collecting all the terms from above, we find that the scalar masses receive additional Hubble-scale mass terms:
\begin{align}
 V_{\rm SUGRA}^{(\text{additive})} \, = \, \sum\limits_i(c_i + d_i \lvert \Phi \rvert^2 + ...)\Lambda^4 \lvert Y_i \rvert^2 + \mathcal{O}(\xi^3), \label{eq:sugraAdditive}
\end{align}
where $c_i$, $d_i \lesssim \mathcal{O}(1)$ are functions of the $\kappa_{ij}$ and $\kappa_{ijk}$. For $\Phi \neq i \neq S$:
\begin{subequations}
\begin{align}
 c_S \, &= \, -4\kappa_{SS},\\
 c_i \, &= \, 1-\kappa_{iS},\\
 d_S \, &= \, 1 - 4\kappa_{SS} - 2\kappa_{\Phi S} - 4\kappa_{\Phi S S} + 12\kappa_{SS}\kappa_{\Phi S} + \kappa_{\Phi S}^2,\\
 d_i \, &= \, 1 - \kappa_{iS} - \kappa_{\Phi S} - \kappa_{i\Phi S} + \kappa_{iS}\kappa_{i\Phi} + 2\kappa_{iS}\kappa_{\Phi S}.
\end{align}
\end{subequations}

The exponential, the bracket in eq.~\eqref{eq:sugraRescaling1}, and the canonical normalization $Y_i \rightarrow Y_i/\sqrt{K_{\overline{i}i}}$ also have the effect of stretching the renormalizable masses $m_{\rm ren}^2$ by a common factor:
\begin{align}
 \frac{ m^2_{i} }{ m^{2}_{i,\rm ren} }  \, =: \, \mathcal{N}_i(\phi) \, = \, \left\{
 \begin{matrix} e^K \left( K_{\overline{i}i} \right)^{-2},   \\
 e^K \left( K_{\overline{i}i} K_{\overline{j}j} \right)^{-1},
 \end{matrix} \right.
 ~~~~\begin{matrix*}[l] \text{for }W \sim Y_i^2,  \\
 \text{for }W \sim Y_i Y_j,~~ j \neq i \text{ fixed}.
 \end{matrix*}
 \label{eq:sugraMultiplicative}
\end{align}
This multiplicative rescaling is mostly irrelevant (except for the one-loop correction due to $A_2$, where we keep it explicitly), so throughout most of our paper, we only include the additive supergravity corrections from eq.~\eqref{eq:sugraAdditive}.

\subsection{Supergravity corrections to fermion masses}
The fermion masses in supergravity can be calculated from
\begin{align}
 \left( m_F \right)_{ij} \, = \, e^{K/2}\left[  W_{ij} + \left( K_{ij} + K_i K_j \right) W + K_i W_j + K_j W_i - K^{k\overline{l}} K_{ij\overline{l}} D_k \right].
\end{align}
During inflation, we can drop all terms proportional to $A_i$, $B_i$, $H$, $N$, $S = \mathcal{O}(\xi)$. This means that $W = 0$, $K_{i\neq \Phi} = 0$, $W_{i\neq S}=0$, $K^{S\overline{l}} = K^{S\overline{S}} \delta_{lS}$:
\begin{align}
 \left( m_F \right)_{ij} \, = \, e^{K/2}\left[  W_{ij} + K_\Phi W_S\left( \delta_{i \Phi}\delta_{j S}+\delta_{j \Phi}\delta_{i S} \right) - K^{S\overline{S}} K_{ij\overline{S}} W_S \right].
\end{align}
The only non-zero terms in $K_{ij \overline{S}}$ are the mixed $\Phi$-$S$ terms:
\begin{align}
 K_{\Phi S\overline{S}} \, = \, K_{S \Phi \overline{S}} \, = \, \Phi^\dagger\left( \kappa_{S\Phi} + 2\kappa_{S\Phi\Phi}\left| \Phi \right|^2 + ...\right),
\end{align}
and therefore the fermionic mass matrix including supergravity corrections is
\begin{align}
 \left( m_F \right)_{ij} \, = \, e^{K/2} W_{ij} + \left( \Delta m_F \right)_{ij},
\end{align}
where the supergravity correction has only one non-vanishing entry during inflation:
\begin{align}
 \left( \Delta m_F \right)_{S\Phi} \, &= \, \left( \Delta m_F \right)_{\Phi S} \, = \, -e^{K/2} W_S \left( K^{S\overline{S}} K_{\Phi S\overline{S} } - K_\Phi \right) \, \notag\\
 &= \, e^{K/2} \Lambda^2 \Phi^\dagger \left(  \kappa_{S\Phi} - 1 \right) \, + \, \mathcal{O}(\Phi^3).
\end{align}
$\Phi$ and $S$ do not have other mixing terms in the fermionic mass matrix, so we can diagonalize the $\Phi$-$S$ block separately. Up to subdominant terms of $\mathcal{O}(\phi^3)$, it takes the simple form
\begin{align}
 m_F \, = \, \left( \kappa_{S\Phi} - 1 \right)\Lambda^2 \Phi^\dagger \begin{pmatrix}
0 & \displaystyle 1 \\
1 & 0 \end{pmatrix},
\end{align}
which leads to the eigenvalues
\begin{align}
 \left| m_{S\Phi}^{(F)} \right| \, = \, \Lambda^2 \left| (1-\kappa_{S\Phi}) \Phi \right|.
\end{align}
With $\Phi \ll \mpl$, this fermion mass is much smaller than the Hubble scale $\Hubble = \Lambda^2/\sqrt{3}$, and the $S$-$\Phi$ fermions remain light during inflation. In particular, their contribution to the one-loop potential is negligible and can be safely ignored in section~\ref{sec:loop}.

The fermion masses also receive a multiplicative stretching factor given by eq.~\eqref{eq:sugraMultiplicative} from the $e^{K/2}$ prefactor and canonical normalization. This rescaling is identical for the scalars and fermions,\footnote{For $\Lambda = 0$, there is no SUSY breaking F-term. As we know that scalar and fermion masses are identical in unbroken SUSY, the mass correction must be identical for scalars and fermions in the limit $\Lambda \rightarrow 0$. The rescaling in eq.~\eqref{eq:sugraMultiplicative}, which does not depend on $\Lambda$, must therefore rescale fermions and scalars in the same way.} and therefore does not affect the cancellation between scalar and fermionic contributions to the one-loop potential in section~\ref{sec:loop}.

\end{document}